\documentclass[aps,prc,groupedaddress,showpacs,showkeys]{revtex4}

\usepackage{hyperref}
\usepackage{amsmath,amssymb,amsbsy}
\usepackage{graphicx}
\usepackage[usenames]{color}

\newcommand{\del}[1]{}
\newcommand{\new}[1]{{#1}}
\newcommand{\newla}[1]{#1}

\newcommand{\fig}[1]{fig.~#1}

\newcommand{\BG}{\text{\begin{tiny}BG\end{tiny}}}

\newcommand{\E}{\mathcal{P}}
\newcommand{\A}{\mathcal{A}}
\newcommand{\Order}{\mathcal{O}}

\newcommand{\MeV}{\; \mathrm{MeV} }
\newcommand{\twopi}{\left( 2\pi \right) }
\newcommand{\GeV}{\; \mathrm{GeV} }
\newcommand{\ekin}{E_{\mathrm{kin}}}
\newcommand{\fm}{\; \mathrm{fm} }

\newcommand{\BUU}{GiBUU }

\begin{document}


\title{Pion induced double charge exchange reactions in the $\Delta$ resonance region
} 


\author{O. Buss}
\email{oliver.buss@theo.physik.uni-giessen.de}
\homepage{http://theorie.physik.uni-giessen.de/~oliver}
\thanks{Work supported by DFG.}
\author{L. Alvarez-Ruso}
\author{A.~B. Larionov}
\author{U. Mosel}
\affiliation{Institut f\"ur Theoretische Physik, Universit\"at Giessen, Germany}
\date{\today}
\begin{abstract}
We have applied the Giessen BUU (GiBUU) transport model to the description of 
the double charge exchange (DCX) reaction \new{of pions with different nuclear targets at incident kinetic energies of $120-180$~MeV}. The DCX process  is highly sensitive to details of the interactions of pions with the nuclear medium and, therefore, represents a major benchmark for any model of \newla{pion} scattering off nuclei at low and intermediate energies. The impact of surface effects, such as the neutron skins of heavy nuclei, is investigated. The dependence of the total cross section on the nuclear mass number is also discussed. We achieve \newla{a good} quantitative agreement with the extensive data set obtained at LAMPF. Furthermore, we compare the solutions of the transport equations obtained in the test-particle ansatz using two different schemes - the full and the parallel ensemble method. 
\end{abstract}

\maketitle

\section{Introduction and Motivation}\label{intro}
Over the last two decades, the Giessen Boltzmann-Uehling-Uhlenbeck (GiBUU) transport model has been developed to describe heavy ion collisions, photon-, electron-, pion- and neutrino-induced reactions within one unified transport framework. A realistic treatment of pion interactions with the nuclear medium is crucial for the interpretation of experiments where pions are produced inside nuclei. Recently, it has been shown using \new{Gi}BUU~\cite{Muhlich:2004zj,Buss:2005bm,Buss:2006vh} that pion rescattering in the final state description of photon induced double-pion production produce a considerable modification of the $\pi\pi$  invariant mass distributions observed by the TAPS collaboration~\cite{Messch}. Moreover, neutrino-induced pion production, a source of background for neutrino oscillation experiments~\cite{Drakoulakos:2004gn}, is very sensitive to pion final state interactions~\cite{Leitner:2006ww,Leitner:2006sp}. The propagation of low-energy pions in nuclear matter in the GiBUU framework has already been extensively discussed in ref.~\cite{Buss:2006vh} and compared to quantum mechanical calculations.

In this context, pionic double charge exchange (DCX) is a very interesting reaction. The fact that DCX requires at least two nucleons to take place makes it a very sensitive \new{benchmark for} pion rescattering and absorption. This reaction received a considerable attention in the past (see for instance Ref.~\cite{LAMPF} and references therein). The mechanism of two sequential single charge exchanges has traditionally been able to explain the main features of this reaction~\cite{Becker:1970tk,Gibbs:1977yz} at low energies although the contribution of the $A(\pi,\pi\pi)X$ reaction becomes progressively important as the energy increases~\cite{Vicente:1988iv,Alqadi:2001pe}. At higher ($\sim1~\mathrm{GeV}$) energies, the sequential mechanism becomes insufficient to account for the reaction cross section~\cite{Abramov:2002nz,Krutenkova:2005nh}.
Extensive experimental studies performed at LAMPF obtained high precision data for doubly differential cross sections on $^3He$~\cite{Yuly:1997ja} and heavier nuclei ($^{16}O$, $^{40}Ca$, $^{208}Pb$)~\cite{Wood:1992bi} in the region of $\ekin=120-270 \MeV$. 

H\"ufner and Thies~\cite{huefnerThies} explored for the first time the applicability of the Boltzmann equation in $\pi N$ collisions and achieved qualitative agreement with data on single and double charge exchange. Their method to solve the Boltzmann equation was based upon an expansion of the pion one-body distribution function in the number of collisions. There, in contrast to our work, the Boltzmann equation is not solved with a test-particle ansatz but by reformulating it into a set of coupled differential equations which can then be solved in an iterative manner. However, this approach was based on simplifying assumptions of averaged cross sections and averaged potentials.
The work by Vicente et al.~\cite{Vicente:1988iv} was based upon the cascade model described in~\cite{osetSimulation}. There, a microscopic model for $\pi N$ scattering was used as input for the pion reaction rates in the simulation. In that work~\cite{Vicente:1988iv}, pion DCX off $^{16}O$ and $^{40}Ca$ has been explored and fair quantitative agreement with data has been achieved.

In our work, we explore DCX on heavier nuclei, comparing with the data measured by Wood et al.~\cite{Wood:1992bi}. We also address the scaling of the total cross section discussed by Gram et al.~\cite{Gram:1989qh}. To focus only on single-pion rescattering, we consider incoming pion energies below $\ekin=180\MeV$; above that energy $2\pi$ production becomes prominent and DCX does not happen necessarily in a two-step process anymore. 
Due to the small mean free path of the incoming pions, the process is mostly sensitive to the surface of the nucleus.
\newla{Therefore, we will discuss and compare two widely used numerical schemes for the solution of the Boltzmann equation: {\it parallel ensemble method} employed in the BUU models \cite{AB85,BBCM86,BertschGupta,Cassing:1990dr} and in the Vlasov-Uehling-Uhlenbeck model \cite{MS85}; and {\it full ensemble method} used in the Landau-Vlasov \cite{Gre87}, Boltzmann-Nordheim-Vlasov \cite{BBD89,BGM94} and Relativistic BUU \cite{FGW96,GFW05} models. 
Both schemes are based on the test-particle representation of the single-particle phase space density, but they differ in the locality of the scattering processes (c.f. discussion in \cite{PhysRevC.40.2611,langBabovsky}). In the {\it parallel ensemble method}, all test-particles are subdivided into the groups, or parallel ensembles. The number of test-particles in each parallel ensemble is equal to the number of physical particles in the system. Collisions are allowed only between test-particles from the same parallel 
ensemble, while the mean field is averaged over all parallel ensembles. Without mean field, thus, the {\it parallel ensemble method} is equivalent to the intranuclear-cascade simulation \cite{Cugnon:1980rb}. In the {\it full ensemble method}, collisions between all test-particles are allowed. The low-energy ($E_{lab}=10-50$ MeV/nucleon) heavy ion collisions are better described by the {\it full ensemble method}, since this method provides convergence to the exact solution of the original kinetic equation in the limit of large $N\equiv$(number of test-particles per nucleon).
However, in codes oriented to the particle production in high-energy heavy ion and hadron-nucleus collisions, the {\it parallel ensemble method} is commonly used, since it is simpler and numerically 
less expensive (see discussion in Sect. III C below). The advantage of the 
{\it parallel ensemble method} is that each parallel ensemble can be considered as a physical event, but this method does not converge to the solution of the kinetic equation in the limit of large $N$.}

This article is structured in the following way. First we introduce our GiBUU~\cite{GiBUUWebpage} transport model emphasizing the most relevant issues. Next, we discuss the solution of the Boltzmann equation using both the full ensemble and the parallel ensemble scheme and check their consistency. Finally, we present our results on DCX in comparison to the data.

\section{The \BUU transport model}\label{buu}
Boltzmann-Uehling-Uhlenbeck (BUU) transport models are based on the Boltzmann equation, which was modified  by Nordheim, Uehling and Uhlenbeck to incorporate quantum statistics. A brief description of the formalism is given below. \new{For a detailed discussion concerning the physical input for pion-induced reactions we refer the reader to~\cite{Buss:2006vh} and the references therein. A more general description of the whole GiBUU model will be given in a forthcoming paper~\cite{GiBUUPaper}.}

\subsection{The BUU equation}\label{buuEQ}
The BUU equation actually consists of a series of coupled differential equations, which describe the time evolution of the \newla{single}-particle \newla{phase-space densities $f_a(\vec{r},\vec{p},t)$. The index $a=\pi,\omega,N,\Delta,\ldots$ denotes the different particle species in our model.} A large number of mesonic and baryonic states is actually included, but at the energies of interest for this study, the relevant ones are $\pi,N$ and the $\Delta(1232)$ resonance.

For a particle of species X, its time evolution is given by
\begin{eqnarray}
\frac{d f_X(\vec{r},\vec{p},t)}{d t}&=&\frac{\partial f_X(\vec{r},\vec{p},t)}{\partial t} +\frac{\partial H_X}{\partial \vec{p}}\frac{\partial f_X(\vec{r},\vec{p},t)}{\partial \vec{r}} 
-\frac{\partial H_X}{\partial \vec{r}}\frac{\partial f_{X}(\vec{r},\vec{p},t)}{\partial \vec{p}}  \nonumber \\  &=&I_{coll}\left(f_X,f_a,f_b,\ldots\right)
\label{BUUEquation}
\end{eqnarray}
with the one-body Hamilton function 
\begin{eqnarray}
H_X(\vec{r},\vec{p})&=&\sqrt{\left(\vec{p}+\vec{A}_X(\vec{r},\vec{p})\right)^{2}+m_X^{2}+U_X(\vec{r},\vec{p})}\nonumber +A_X^{0}(\vec{r},\vec{p}) \; .
\label{HamiltonFunc}
\end{eqnarray}
The scalar potential $U_X$ and the vector potential $A^\mu_X$ of species X may in principle depend upon the phase space densities of all other species. Hence, the differential equations are already coupled through the mean fields. In the limit of $I_{coll}=0$, eq.~(\ref{BUUEquation}) becomes the well-known Vlasov equation. The collision term $I_{coll}$ on the right-hand side of eq.~(\ref{BUUEquation}) incorporates explicitly all scattering processes among the particles. The  reaction probabilities used in this collision term are chosen to match the elementary collisions among the particles in vacuum. 
Within the BUU framework \newla{the $\pi N$ reaction cross section is given by an incoherent sum of resonance contributions and a direct, i.e. point-like, contribution.} Interference effects are therefore neglected. 

\subsection{Elementary processes}
\begin{figure}[b]
\begin{center}
\includegraphics[]{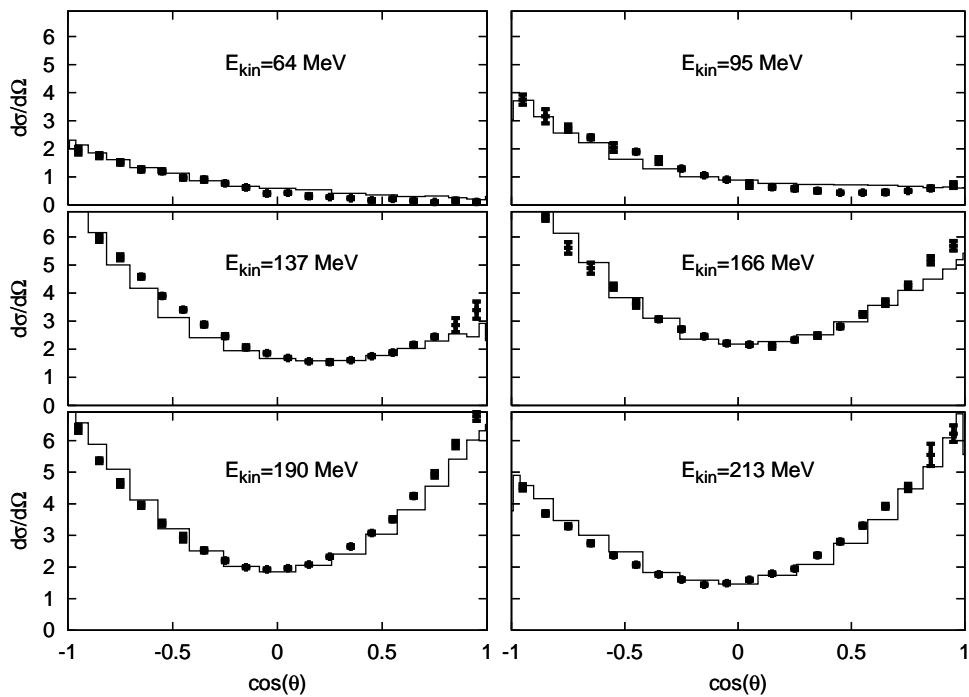}
\end{center}
\caption{The angular distributions for the charge exchange process $\pi^- p \to \pi^0 n$ in the CM-frame of pion and proton. The plots are labeled by the kinetic energies of the pions in the laboratory frame. The data are taken from ref. \cite{Sadler:2004yq}.}
\label{eleX}
\end{figure}

DCX emerges from the interplay of two elementary mechanisms: pion-nucleon quasielastic scattering, with or without charge exchange, and pion absorption inside the nucleus. In our model, the cross section for quasi-elastic scattering is given by an incoherent sum of background $\sigma^{\BG}$ and resonance contributions
\begin{eqnarray}
\sigma_{\pi N \rightarrow \pi N}&=&\sigma_{\pi N \rightarrow R \rightarrow \pi N}+\sigma^{\BG}_{\pi N \rightarrow \pi N} \; . \nonumber \label{backDef}
\end{eqnarray}
The resonance cross sections are obtained from the partial wave analysis of ref.~\cite{ManleySaleski}. The background cross sections denoted by $\sigma^{\BG}$ are chosen in such a manner, that the elementary cross section data in the vacuum are reproduced. Background contributions are instantaneous in space-time, whereas the resonances propagate along their classical trajectories until they decay or interact with one or two nucleons in the medium. 
 
As an improvement in the earlier treatment \cite{Buss:2006vh,Effenberger:1996rc,Engel:1993jh}, we now included a more realistic angular distribution for the elastic scattering of the pions. Due to the $P$-wave nature of the $\Delta(1232)$ resonance, we assume for $\pi N \to \Delta \to \pi N$ in the \new{resonance} rest frame a distribution of the pion scattering angle $\theta$ according to 
\[ f^\Delta(s,\theta)=\left(1+3\cos^2(\theta) \right) g(s,\theta) \]
which is peaked in forward and backward scattering angles. The function $g(s,\theta)$, depending on Mandelstam $s$, parameterizes the energy dependence of the $\pi N$ angular distribution. In a coherent calculation the angular distribution is generated by interference effects, which can not be accomplished by our transport model. In our ansatz we need to split the cross section in an incoherent way to preserve our semi classical resonance picture. Therefore we take
\[
g(s,\theta)=\left(\alpha-\cos(\theta) \right)^{\beta \left(m_\Delta-\sqrt{s}\right)/m_\Delta}
\]
with the $\Delta$ pole mass $m_\Delta=1.232 \GeV$. For the background events we assume  
\[ f^\BG(s,\theta)=g(s,\theta) \; . \]
The constants $\alpha=1.9$ and $\beta=26.5$ are fitted to the angular distributions measured by Crystal Ball~\cite{Sadler:2004yq}; a comparison of our parameterization to this data is shown in \fig{\ref{eleX}}.

Concerning pion absorption, the most important mechanisms are a two step processes in which $\pi N \to \Delta$ is followed by $\Delta N \to NN$ \new{or $\Delta N N \to NNN$}, and a one step background process $\pi N N \to N N$. 
As medium modifications, we include Pauli-blocking and Fermi-motion of the nucleons, Coulomb forces and hadronic potentials for baryons; we account as well for the collisional broadening in the $\Delta$ resonance width. \new{In \cite{Buss:2006vh} we discussed the influence of the real part of the pion self energy on absorption processes in nuclei. Since DCX is mostly sensitive to the surface (see the discussion in sec. \ref{resu}), the influence of this real part has turned out to be negligible due to the low effective density and has not been included in the calculations presented here.}
For details see ref. \cite{Buss:2006vh} and the references therein.

\subsection{Nuclear densities} \label{densSec}
The nucleons are initialized in a local density approximation. 
For $^{16}_{8}O$, $^{40}_{20}Ca$, $^{208}_{82}Pb$ we have implemented density profiles $\rho(r)$ according to the parameterizations collected in ref.~\cite{osetPionicAtoms}, which are of Woods-Saxon type for heavier nuclei ($Ca$, $Pb$) and of harmonic-oscillator type for lighter ($O$) ones. The proton densities are based on the compilation of ref.~\cite{DeJager:1974dg} from electron \newla{scattering. The neutron densities are provided by Hartree-Fock calculations. 
For $^{103}_{45}Rh$, we use a Woods-Saxon density distribution
\begin{eqnarray}
\rho_{n}(r)&=&  \frac{\rho_n^0}{1+exp((r-R_n)/a_n)} \nonumber \\
\rho_{p}(r)&=&  \frac{\rho_p^0}{1+exp((r-R_p)/a_p)}\; ,
\label{densWS}
\end{eqnarray}
with the parameters given in table I~\cite{lenske}. }
\begin{table}
\begin{center}
\begin{tabular}{cc|cc|cc}
 $\rho_p^0 [fm^{-3}]$& $\rho_n^0 [fm^{-3}]$ & $R_p [fm]$ & $R_n [fm]$ & $a_p [fm]$ &$a_n [fm]$\\ 
\hline
 0.0708 & 0.0835  &  5.194 & 5.358 & 0.4743 & 0.4780
\label{WS_Rh_table}
\end{tabular}
\caption{Parameters of the Woods-Saxon parametrizations for $_{45}^{103} Rh$.}
\end{center}
\end{table}
The larger neutron radii of heavy nuclei play a relevant role in DCX as we show in Sec.~\ref{resu}.

\section{Solution of the Boltzmann equation}\label{parallelFull}

The fact that the DCX reaction depends considerably on the \new{spatial} distributions of protons and neutrons implies that it is also sensitive to the degree of locality of the scattering processes. In the non-discretized version of the BUU equation, the interactions are strictly local in space-time. Utilizing the so called \textit{test-particle ansatz} to solve the problem numerically, this is no longer the case. Therefore, \new{we elaborate} in this section on this degree of locality of the scattering processes in our simulation. As a first step, \new{we point} out the connection of the underlying BUU equation to the actual numerical implementation. This will lead us to a proper definition of a typical volume in which particles are allowed to interact. Hereafter, we will introduce an approximation to the full solution: \textit{The parallel ensemble scheme}. In this scheme computation time is radically decreased by reducing the locality. It is widely employed to solve the BUU equation in particle physics (e. g.~\cite{AB85,BBCM86,BertschGupta,Cassing:1990dr,Effenberger:1996rc}) since otherwise computations would not be feasible. In  sec. \ref{resu}, we will discuss a possible problem related with this scheme and evaluate its applicability to the DCX process.

\subsection{The collision term}
For the sake of simplicity, let us consider a model with only one fermionic particle species which has no degeneracy, allowing only for binary scattering processes
\begin{eqnarray*}
A(\vec{p_{A}})\ B(\vec{p_{B}}) \longrightarrow a(\vec{p_{a}}) \ b(\vec{p_{b}}) \; .
\end{eqnarray*}
If such a scattering event occurs at point $\vec{r}$, then the \newla{single-particle phase-space density} decreases in the vicinity of phase space points $(\vec{r},\vec{p_{A}})$ and $(\vec{r},\vec{p_{B}})$ and increases in the vicinity of $(\vec{r},\vec{p_{a}})$ and $(\vec{r},\vec{p_{b}})$.
Therefore, at each phase space point, the collision term in eq.~(\ref{BUUEquation}) consists of a gain term due to particles which are scattered into this phase space point and a loss term due to particles which are scattered out:
\begin{eqnarray*}
\frac{d f(\vec{r},\vec{p_{A}},t)}{d t}\nonumber &=&I_{\mathrm{gain}}(\vec{r},\vec{p_{A}},t)-I_{\mathrm{loss}}(\vec{r},\vec{p_{A}},t)
\end{eqnarray*}
with   
\begin{eqnarray*}
I_{\mathrm{loss}}(\vec{r},\vec{p_{A}},t)&=&\frac{1}{(2\pi)^3}\int d^{3}p_{B}\int  
d\Omega_{CM}
\frac{d\sigma_{AB\to ab}}{d \Omega_{CM}} 
v_{AB} \ 
f(\vec{r},\vec{p_{A}},t)\ f(\vec{r},\vec{p_{B}},t) \E_{a} \E_{b} \; ,\\
I_{\mathrm{gain}}(\vec{r},\vec{p_{A}},t)&=&\frac{1}{(2\pi)^3}\int d^{3}p_{a}\ \frac{1}{(2\pi)^3}\int d^{3}p_{b} 
(2\pi)^3  \frac{d\sigma_{ab\to AB}}{d^3p_A}   v_{ab} \ 
f(\vec{r},\vec{p_{a}},t)\ f(\vec{r},\vec{p_{b}},t) \E_{A} \E_{B} \; ,\\
\end{eqnarray*}
\newla{where $d\sigma_{AB\to ab}/\Omega_{CM}$ and $d\sigma_{ab\to AB}/d^3p_A$ are the angular and momentum differential cross sections for the reactions $AB\to ab$ and $ab\to AB$, respectively. $v_{AB}$ and $v_{ab}$ are the relative velocities of the collision partners $AB$ and $ab$. The terms $\E_X=1- f(\vec{r},\vec{p_{X}},t)$ with $X=A, B, a,b$ correspond to the Pauli blocking of the final states. The momenta of the particles in $I_{\mathrm{loss}}$ and $I_{\mathrm{gain}}$ satisfy the condition of energy and momentum conservation.}

To point out the connection between our numerical implementation and the underlying BUU equation, \new{we concentrate} on the loss term of BUU. We will therefore not elaborate on the gain term $I_{\mathrm{gain}}$ which describes the production of particles. However, its numerical implementation is analogous to the loss term since both are connected by detailed balance.


The Boltzmann equation can be solved numerically using the so called test-particle ansatz where the single-particle density is expressed in terms of $\delta$-functions 
\[
f(\vec{r},\vec{p},t)=\lim_{N\to\infty} \frac{\twopi^3}{N} \sum_{i=1}^{\A \times N} \delta(\vec{r}-\vec{r}_i(t))\delta(\vec{p}-\vec{p}_i(t)) \; .
\]
\newla{Here $\A$ denotes the number of physical particles and $N$ is the number of test-particles per physical one. So the single-particle phase-space density is interpreted as a sum of all test particle densities.}
\newla{The centroids of the $\delta$-functions $\vec{r}_i$ and $ \vec{p}_i$  obey the classical Hamiltonian equations. The change of the single-particle phase-space density per infinitesimal $\Delta t$ due to collisions is given by}
\begin{eqnarray}
\Delta  f(\vec{r},\vec{p},t) &=& \Delta t \left( I_{\mathrm{gain}}-I_{\mathrm{loss}} \right) \; .
\end{eqnarray}
In terms of the test-particle ansatz the loss term reads
\begin{eqnarray}
\Delta t \ I_{\mathrm{loss}}(\vec{r},\vec{p_{A}},t)
%
&=& \lim_{N\to\infty} \frac{\twopi^3}{N} \sum_{i=1}^{\A\times N} \sum_
{\begin{array}{l}  j=1 \\    j\neq i  \end{array}}^{\A\times N} \delta(\vec{p}_A-\vec{p}_i)  \delta(\vec{r}-\vec{r}_i) \lim_{N\to\infty} \underbrace{\frac{1}{\sigma_{ij}}\int d\Omega_{CM} \; \E_{a} \E_{b} \frac{d\sigma_{ij\to ab}}{d \Omega_{CM}}}_{=\overline{\E_a\E_b}} \nonumber\\ 
&& \times \underbrace{   \sigma_{ij}  \Delta t \, v_{ij} \frac{1}{N}}_{=\Delta V_{ij}}   \delta(\vec{r}-\vec{r}_j) \\ 
&=& \lim_{N\to\infty} \left( \frac{\twopi^3}{N} \sum_{i=1}^{\A\times N} \sum_{\begin{array}{l} j=1 \\  j\neq i  \end{array}}^{\A \times N}  \delta(\vec{p}_A-\vec{p}_i)  \delta(\vec{r}-\vec{r}_i) \overline{\E_a\E_b}
\int_{\Delta V_{ij}}\delta(\vec{r}\; '-\vec{r}_j) d^3r' 
\right) \label{loss}
\end{eqnarray}
\newla{where $\sigma_{ij}$ and  $v_{ij}$ are the total interaction cross section and the relative velocity of the test-particles $i,j$;  $\Delta V_{ij}=\sigma_{ij} \Delta t \, v_{ij} /N$ is an infinitesimal volume in the vicinity of $\vec{r}_i$}. Note, that the latter volume defines the locality of the scattering process of two test-particles. The term $\overline{\E_a\E_b}$ denotes the blocking of the final state averaged over its angular distribution. We excluded self-interactions - therefore a test-particle cannot scatter with itself. 

\subsection{Numerical implementation}
In a real calculation the number of test-particles $N$ is chosen finite, for our purposes usually of the order of 300-1500. \newla{The time step $\Delta t$ is chosen such that the average distance travelled by the particles during $\Delta t$ is less than their mean free path. Therefore, $\Delta V_{ij}$ is small enough such that a particle has no more than one scattering partner at a given time step.} The algorithm proceeds as a sequence of the following steps:
\begin{itemize}
\item First, we propagate at each time step the test-particles according to Hamilton's equations. 
\item The loss term is implemented according to equation~(\ref{loss}). Therefore we consider each term in eq.~(\ref{loss}) separately. For simplicity let us just consider one summand describing the loss of the $i$th test-particle due to a collision with the $j$th.
\begin{enumerate}
\item 
The term $\int_{\Delta V_{ij}}\delta(\vec{r}\; '-\vec{r}_j) d^3r'$ gives $1$ or $0$ depending on the fact whether $j$ is \newla{within $\Delta V_{ij}$. The volume $\Delta V_{ij}$ is  chosen to be a cylinder of height $\Delta t v_{ij}$ with a circle basis $\sigma_{ij}/N$; the symmetry axis is chosen along $\vec{v}_{ij}$ and the basis is centered at $\vec{r}_i$. This corresponds to the usual minimum distance concept \cite{Cugnon:1980rb}.}
\item 
If the result of the integral is $1$, then we evaluate 
\[
\overline{\E_a\E_b}=
 \frac{1} {\sigma_{ij}} \int d\Omega_{CM} \; \E_{a} \E_{b} \frac{d\sigma_{ij\to ab}}{d \Omega_{CM}} \; .
\]
For this we make a Monte-Carlo integration with only one integration point, which is a good approximation in the large $N$ limit. This one point $\Omega^{\mathrm{CM}}$ is chosen in the center of mass (CM) frame randomly according to the weight $\frac{1}{\sigma_{ij}}\frac{d\sigma_{ij\to ab}}{d \Omega_{CM}}$.  Since $\sqrt{s}$ is fixed, $\Omega^{\mathrm{CM}}$ defines the random momentum $\vec{p}_a\,^{\mathrm{CM}}$. \newla{Furthermore,}
 \[
\vec{p}_b\,^{\mathrm{CM}}=-\vec{p}_a\,^{\mathrm{CM}} \; .
\]
\newla{Finally, by boosting the momenta to the computational frame, we get}
\[
\overline{\E_a\E_b}=(1- f(\vec{r},\vec{p}_{a},t))(1- f(\vec{r},\vec{p}_{b},t)) \; .
\]
\item 
Now we interpret $\overline{\E_a\E_b}$ as a probability that the reaction takes place. So we make a second Monte-Carlo decision on whether we accept the reaction or not. This corresponds to substituting $\overline{\E_a\E_b}$ by a Bernoulli distributed random number with $p=\overline{\E_a\E_b}$. With this substitution, the expectation value of the summand is equal to the original summand. In the limit of many ensembles $N$, i.e. many summands, this yields the right loss term. 
\item
If the reaction is accepted, then we get for this event where $i$ is scattering with $j$
the loss contribution
\[
  \delta(\vec{p}_A-\vec{p}_i)  \delta(\vec{r}-\vec{r}_i)
\]
which corresponds to the destruction of the $i$th test-particle. Due to the double sum in eq.~(\ref{loss}), we get also the contribution
\[
  \delta(\vec{p}_A-\vec{p}_j)  \delta(\vec{r}-\vec{r}_j) \; .
\]
This latter term corresponds to the destruction of the $j$th test-particle. Note that we do not evaluate $\overline{\E_a\E_b}$ for this case, but take the same value which lead to the destruction of $i$. This reflects that energy is conserved on an event-by-event basis.
\item
In our simulation, the final states with momenta $\vec{p}_b\,^{\mathrm{CM}}$ and $\vec{p}_a\,^{\mathrm{CM}}$ contribute to the gain term. New test particles with those momenta are therefore added to the simulation.
\end{enumerate}
\end{itemize}

A generalization to $2\to3$ and $3\to2$ processes and to finite particle species including degeneracies is straight-forward.


\subsection{Full and parallel ensemble method}

The kind of simulation we described in the last section, is called a \textit{full ensemble} calculation. There exists a common simplification to this method: \textit{the parallel ensemble method}~\cite{BertschGupta}. In this scheme one sets $N=1$, performs $\tilde{N}$ runs at the same time and then averages the results over all runs. 
The densities used in each run are the averaged densities of all $\tilde{N}$ parallel runs. Therefore the propagation part stays the same, whereas the collision term gets very much simplified. 

Note that the only justification for this simplification is a great gain in computation time. In a full ensemble method, the propagation part scales according to the number of test-particles per nucleon $N$, whereas the collision term scales with $N^2$ - therefore the computation time is $\Order (N^2)$. In a parallel ensemble method $\tilde{N}$ runs are performed, which results in $\Order (\tilde{N})$ computation time. So there is linear scaling in a parallel ensemble run, but a quadratic one in a full ensemble run. \new{In pioneering works, it was shown by Welke et al. \cite{PhysRevC.40.2611} and Lang et al.~\cite{langBabovsky}, that the parallel ensemble scheme is a good approximation to the full ensemble scheme under the conditions of high-energy heavy ion collisions.}

\section{Results of the simulations}\label{resu}

\subsection{Comparison of full and parallel ensemble runs}
In the previous section we introduced the concept of the parallel ensemble approximation. For DCX, surface effects are expected to be important, therefore the spatial resolution could be relevant in this context.
Indeed, a major problem of the parallel ensemble scheme is that the volume $\Delta V_{ij}$ can become very large. In the energy regime under consideration, the incoming pions interact strongly with the nucleons so that the total cross section can reach more than $200 \mbox{ mb}$. For a parallel ensemble run, the typical volume has therefore the size of $5 \fm^3$ for a typical $\Delta t=0.25\fm/c$. Since it is not obvious that the parallel ensemble scheme should be reliable in this regime, we hence decided to evaluate this approximation scheme by comparison to the full ensemble method.

\begin{figure}[h]
\begin{center}
\includegraphics[]{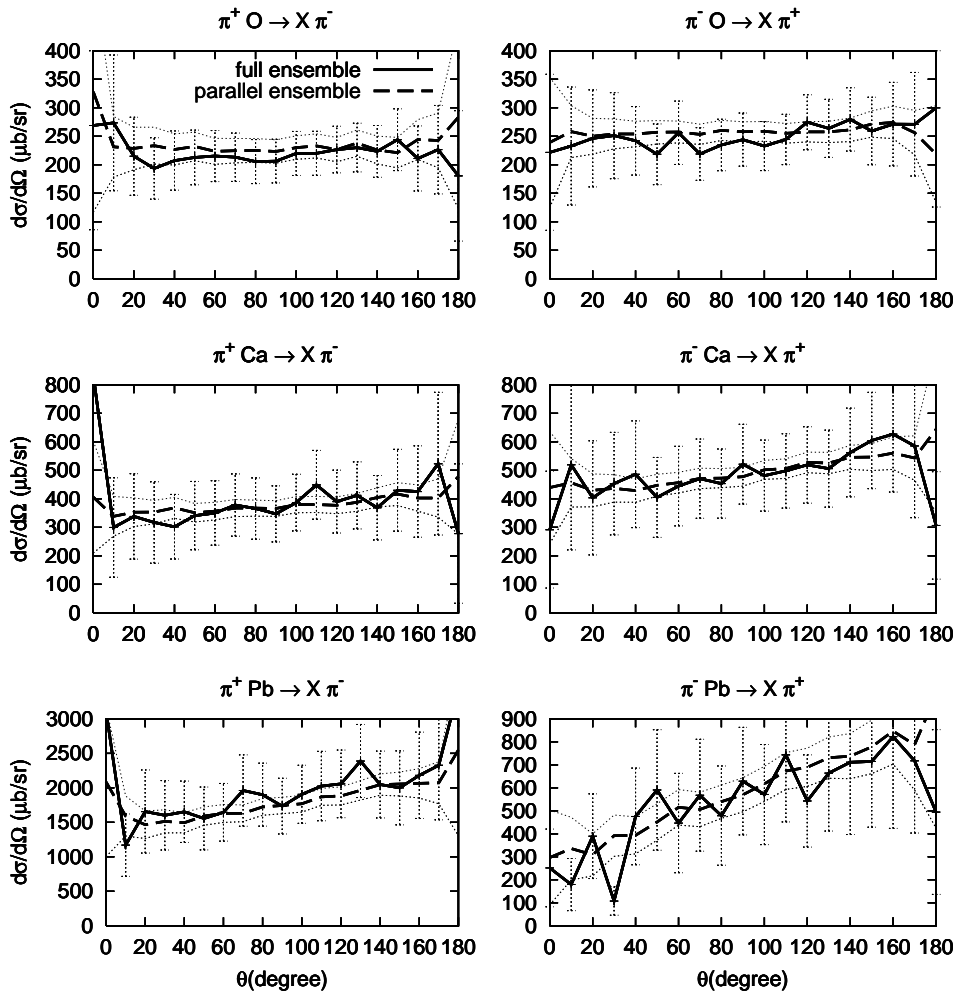}
\end{center} 
\caption{Comparison of the full and parallel ensemble methods. Angular distributions for the double charge exchange process $\pi^\pm A\to \pi^\mp X$ at $E_{\mathrm{kin}}=180 \MeV$. The error bars denote the $1\sigma$ statistical error of each point of the full ensemble run; the thin lines denote the $1\sigma$ statistical error of the parallel result.}
\label{paraFull}
\end{figure}

In \fig{\ref{paraFull}}, the results for $d\sigma /d\Omega$ at $180 \MeV$ kinetic energy of the pion are presented for both methods. The results obtained are consistent with each other, therefore we used the parallel ensemble method to save CPU time and to reach higher statistics for all further calculations. Note that in the present problem, in order to obtain a result at a given energy for one specific nucleus one CPU-day is required in the parallel ensemble scheme. In the full scheme this takes of the order of $30$ CPU-days for an acceptable statistics, as shown in \fig{\ref{paraFull}}.

\subsection{Influence of the density profile} \label{densChapter}
\new{The DCX is, due to the low pion mean free path in nuclear matter~\cite{Buss:2006vh}, very sensitive to the surface properties of the nuclei. Therefore, we compared the results with our present density parameterization for $^{208}Pb$ \cite{osetPionicAtoms}, as described in section \ref{densSec}, to the results obtained with the one used in previous publications~\cite{Buss:2006vh,Muhlich:2004zj,Effenberger:1996rc}. For $^{208}Pb$, both distributions are parametrized according to eq.~(\ref{densWS}) with the parameters given in table II.} 
\begin{table}
\begin{center}
\begin{tabular}{l|cc|cc|c}
 & $\rho_p^0\; [fm^{-3}]$& $\rho_n^0\; [fm^{-3}]$ & $R_p\; [fm]$ & $R_n\; [fm]$ & $a_p=a_n\; [fm]$ \\ 
\hline
default parameterization \cite{osetPionicAtoms} & 0.0631  & 0.0859 & 6.624 & 6.890 &0.549\\ 
old parameterization \cite{Buss:2006vh}         & 0.0590  & 0.0900 & 6.826 & 6.826 & 0.476\label{WStable}
\end{tabular}
\caption{Parameters of the Woods-Saxon parametrizations for $_{82}^{208} Pb$.}
\end{center}
\end{table}
However, neutron skins are very interesting because in those skins only $\pi^+$ mesons can undergo charge exchange reactions. For the positive pions this causes an enhancement of DCX processes at the surface, so the pions do not need to penetrate deeply for this reaction. Hence, the probability for their absorption is reduced. As can be observed in fig. 3, the enhancement in the cross section due to the neutron skin is roughly $35\%$ at $180\MeV$. We conclude that surface effects are prominent and can not be neglected. Note that for the $\pi^-$ similar arguments lead to a reduction of the cross section.


\begin{figure}
\label{densVer}
\begin{center}
\includegraphics[]{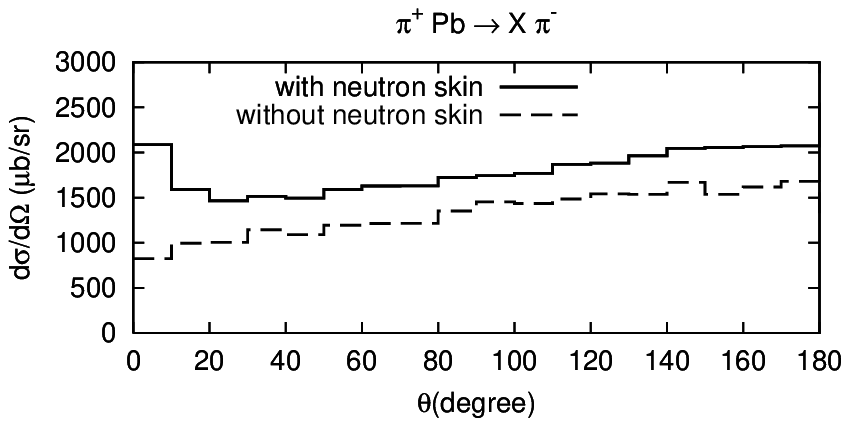}
\caption{Influence of the density distribution on the angular distributions for the double charge exchange process $\pi^+ Pb\to \pi^- X$ at $E_{\mathrm{kin}}=180 \MeV$. \new{The solid line shows the result obtained with our present density distribution \cite{osetPionicAtoms}; the dashed line was obtained with the previously used one~\cite{Buss:2006vh,Muhlich:2004zj,Effenberger:1996rc}, which contains no neutron skin.}}
\end{center}
\end{figure}


\subsection{Comparison to data}
Now we proceed to the comparison with the data measured at LAMPF by Wood et al.~\cite{Wood:1992bi}. We discuss first the total cross section. Hereafter, we will explore angular distributions and, finally, the double differential cross sections as a function of both angles and energies of the outgoing pions are addressed. 

\begin{figure}
\begin{center}
\includegraphics[]{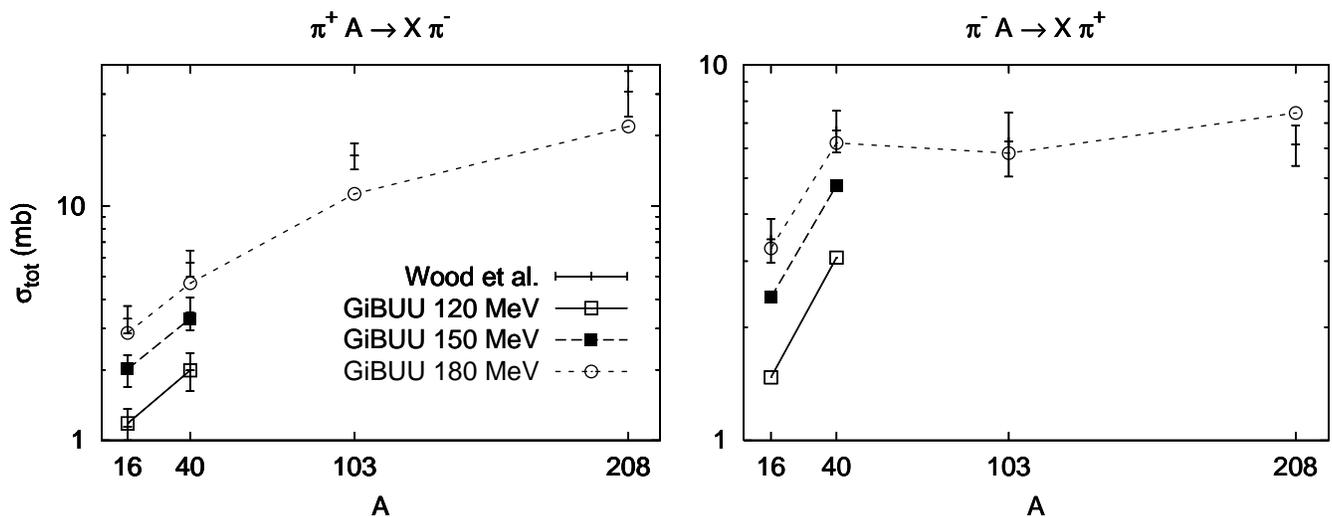}
\end{center}
\caption{\new{The inclusive double charge exchange total cross section as function of the nuclear target mass at $E_{\mathrm{kin}}=120,150$ and $180 \MeV$. The lines connecting our results are meant to guide the eye; the data are taken from Ref.~\cite{Wood:1992bi} (left panel: $E_{\mathrm{kin}}=120,150$ and $180  \MeV$, right panel: only $180 \MeV$).}}
\label{total}
\end{figure}

In \fig{\ref{total}} one can see the excellent quantitative agreement to the total cross \newla{section} data at $120$, $150$ and $180 \MeV$ for Oxygen and Calcium. For the Lead nucleus we see some discrepancies. Notice that we reproduce the different $A$ \new{dependencies} of \new{both} $(\pi^+,\pi^-)$ and $(\pi^-,\pi^+)$ reactions. It is due to the fact that, when $A$ increases, the number of neutrons increases with respect to the number of protons, and this favors the $\pi^+$ induced reaction.

\begin{figure}
\begin{center}
\includegraphics[]{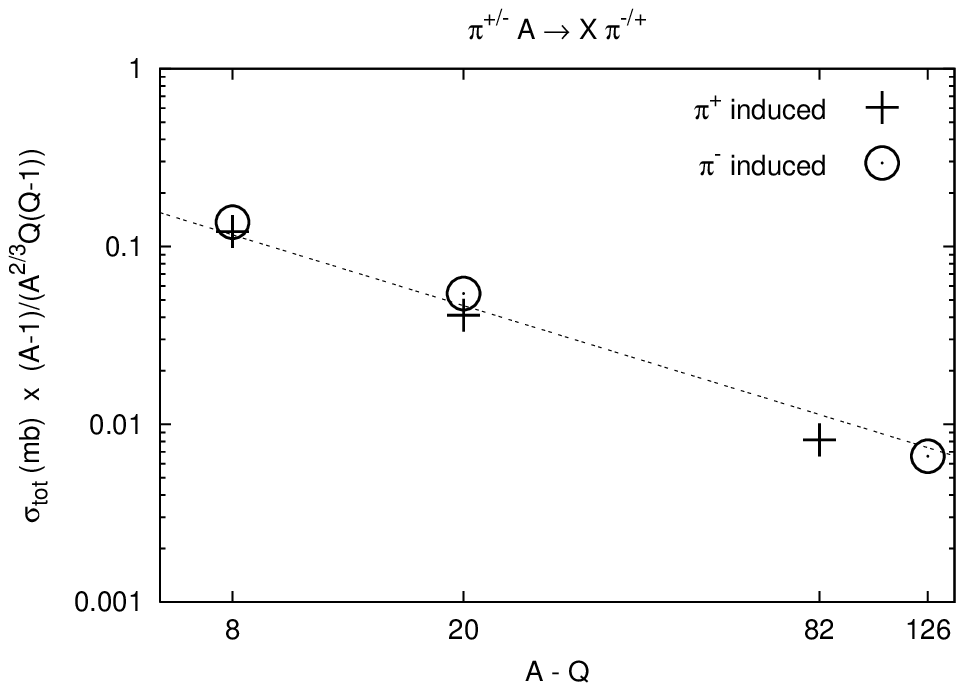}
\end{center}
\caption{The scaling of the total charge exchange cross section according to eq.~(\ref{scalingEQ}) is visualized by dividing $\sigma_{tot}$ by the factor $A^{2/3}Q(Q-1)/(A-1)$  and plotting it as a function of $A-Q$. $Q$ denotes the number of protons in the case of $(\pi^-,\pi^+)$ and the number of neutrons in $(\pi^+,\pi^-)$. The points are GiBUU results at pion kinetic energies of $180 \MeV$; the dashed line
denotes a function proportional $1/(A-Q)$, corresponding to the exact scaling.}
\label{scaling}
\end{figure}

In \cite{Gram:1989qh}, Gram et al. discuss a scaling law of the total cross section. They argue as follows. Since the first collision takes place predominantly at the surface, the cross section should scale with $A^{2/3}$. Furthermore they assume that  DCX is mainly a two-step process, and that a pion which undergoes an elastic process at the first collision will not contribute. This is reasonable because the incoming pions loose energy in the elastic process, and their cross section for a second charge-exchange reaction is hereafter very much reduced. For a negative pion the first charge exchange reaction occurs with a probability of $Z/N$ where $Z(N)$ denotes the number of protons(neutrons). This is the case if the interaction is dominated by the $\Delta$ resonance, as it should be in this energy region. Finally, the second charge exchange process then takes place with the probability $(Z-1)/(A-1)$ since, in the isospin limit, the $\pi^0$ interacts equally well with protons and neutrons.  
Putting these considerations together and extending them to the $\pi^+$ case, the cross section for DCX is expected to scale according to
\begin{eqnarray}
\sigma_{tot}\sim A^{2/3}  \frac{Q}{A-Q}\frac{Q-1}{A-1} \; ,
\label{scalingEQ}
\end{eqnarray}
where $Q$ denotes the number of protons in the case of $\pi^-$ induced and the number of neutrons in $\pi^+$ induced DCX. 

Gram et al. \cite{Gram:1989qh} find good agreement of this scaling law with experimental data. Also in the GiBUU simulation this scaling is fulfilled as can be seen in \fig{\ref{scaling}}. Nevertheless, one may wonder why this scaling law works in a process which is so sensitive to the neutron skin on heavy nuclei, as has been shown in fig. 3. Since the first collision takes place on the surface, a neutron skin causes an enhancement in the $A(\pi^+,\pi^-)X$ reaction while $A(\pi^-,\pi^+)X$ is suppressed. This effect leads to a deviation from the scaling. However there are also Coulomb forces which are not negligible. The Coulomb force enhances $A(\pi^-,\pi^+)X$ by attracting the negative projectiles and repelling the positive products, which therefore have a smaller path in the nucleus and undergo less absorption. And, due to similar arguments, the reaction $A(\pi^+,\pi^-)X$ is suppressed. We find that this effect counteracts the one from the neutron skin restoring the scaling.  In any case\new{,} the approximate scaling exhibited by the cross section shows that the reaction is very much surface driven and can be very well understood in terms of a two-step process.

\begin{figure}
\begin{center}
\includegraphics[]{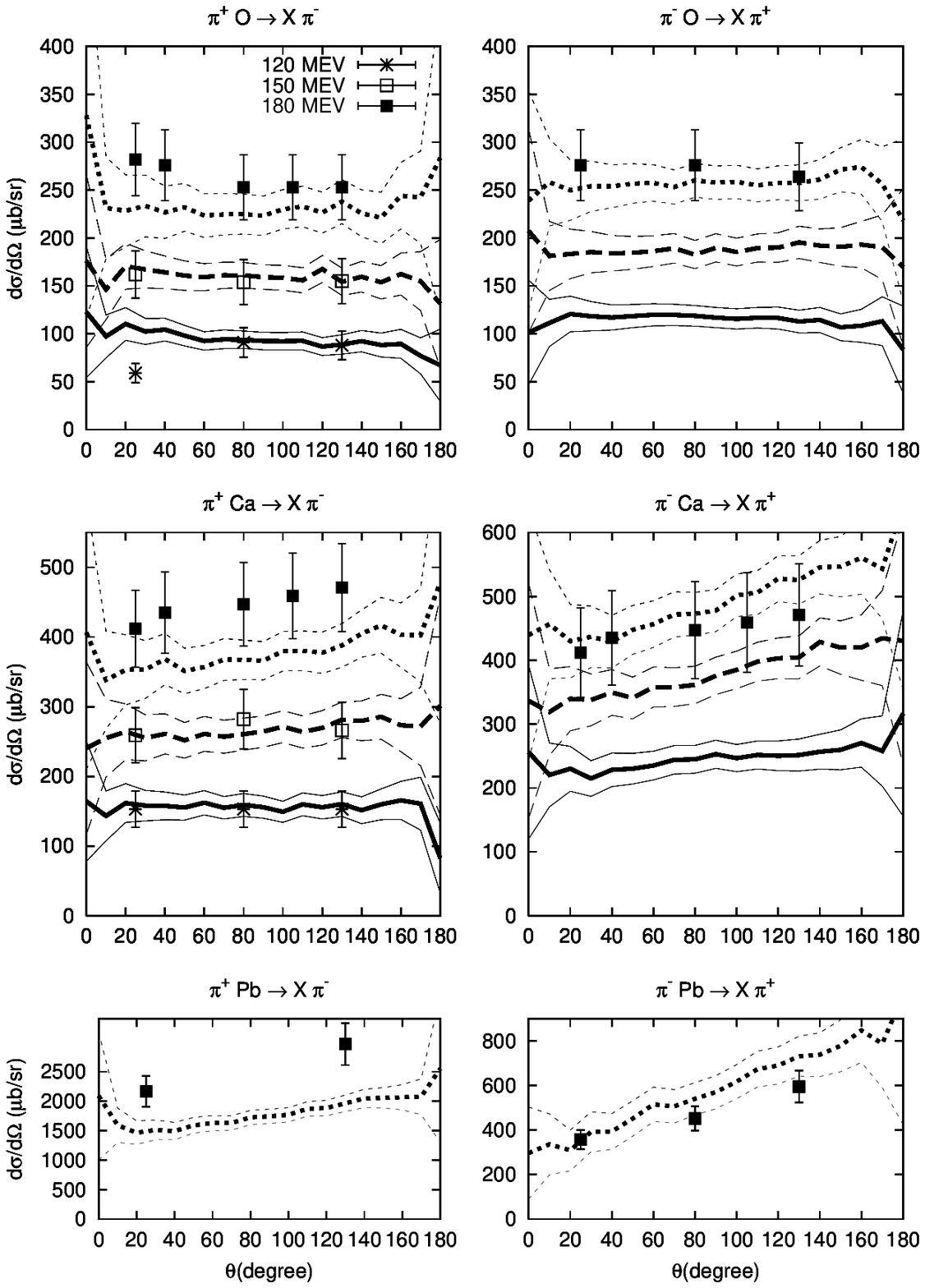}
\end{center}
\caption{Angular distributions for the double charge exchange process $\pi^\pm A\to \pi^\mp X$ at $E_{\mathrm{kin}}=120,150$ and $180 \MeV$. The data points are taken from \cite{Wood:1992bi}; only systematical errors are shown. The solid lines represents the GiBUU results at $120 \MeV$, the dashed lines at $150 \MeV$ and the dotted lines at $180 \MeV$. The bold lines represent the results while the thin lines represent a $1\sigma$ confidence level for each point based upon our statistics. 
}
\label{parallel}
\end{figure}

In \fig{\ref{parallel}} we show $d\sigma / d\Omega$ for DCX at $E_{\mathrm{kin}}=120,150$ and $180 \MeV$ on $^{16}O$, $^{40}Ca$ and $^{208}Pb$ as a function of the scattering  angle $\theta$ in the laboratory frame. Our results 
(bold lines) are shown together with their uncertainties of statistical nature (thin lines). The latter ones are well under control except at very small and very large angles, where statistics is very scarce. Again, there is a very good quantitative agreement for both $O$ and $Ca$. In the $Pb$ case, the $(\pi^-,\pi^+)$ reaction is well described, but the $(\pi^+,\pi^-)$ one is underestimated in spite of the enhancement caused by the neutron skin.  

\begin{figure}
\begin{center}
\includegraphics[]{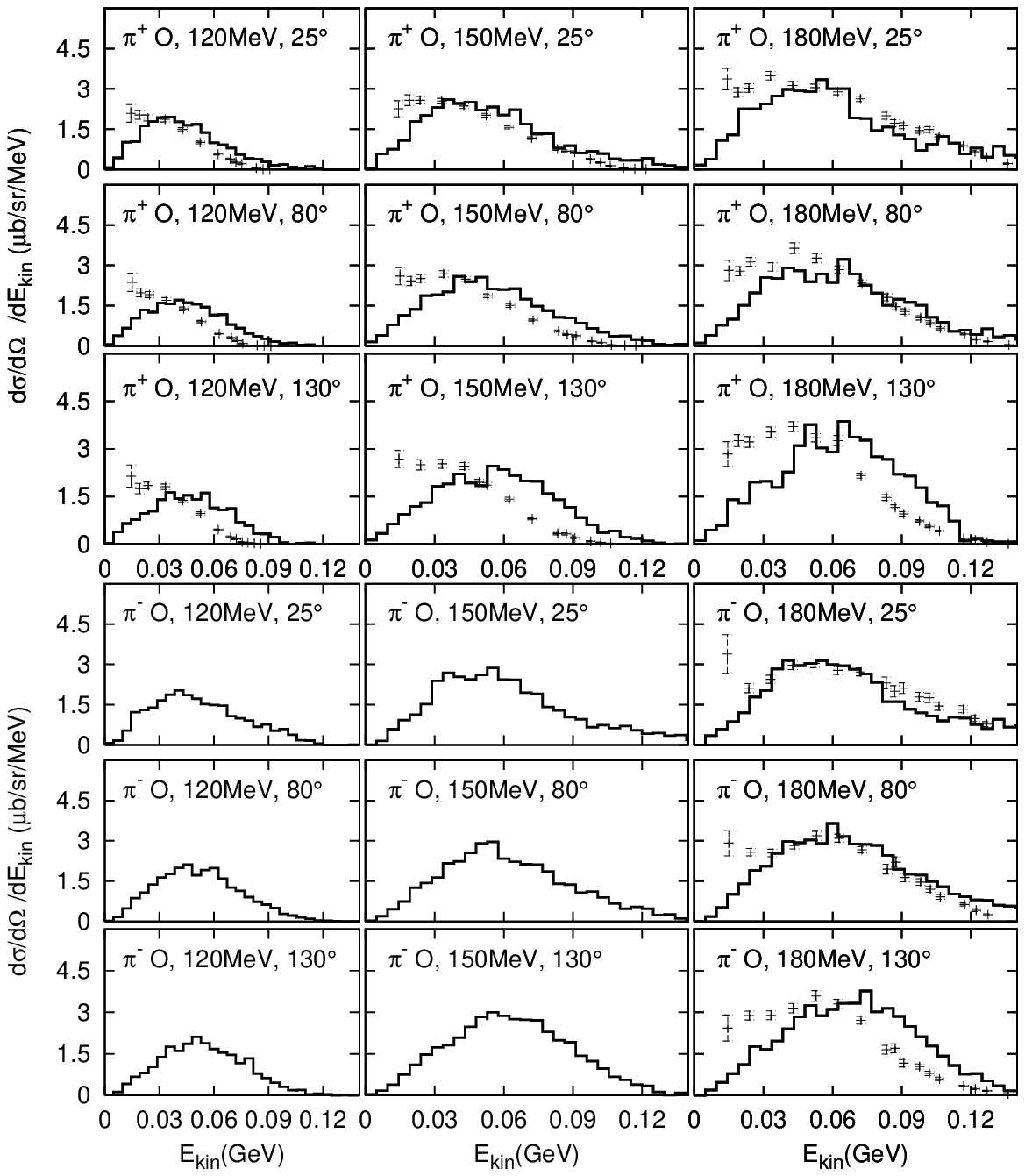}
\end{center}
\caption{Double differential cross sections for the DCX process $\pi^\pm O\to \pi^\mp X$ at $\ekin=120, 150 \mbox{ and } 180 \MeV$. The results at different angles are shown as function of the kinetic energies of the produced pions. Data are taken from \cite{Wood:1992bi}, only statistical errors are shown. The GiBUU results are shown as histograms, where the fluctuations indicate the degree of statistical uncertainty.}
\label{O_dOmegadE}
\end{figure}

\begin{figure}
\begin{center}
\includegraphics[]{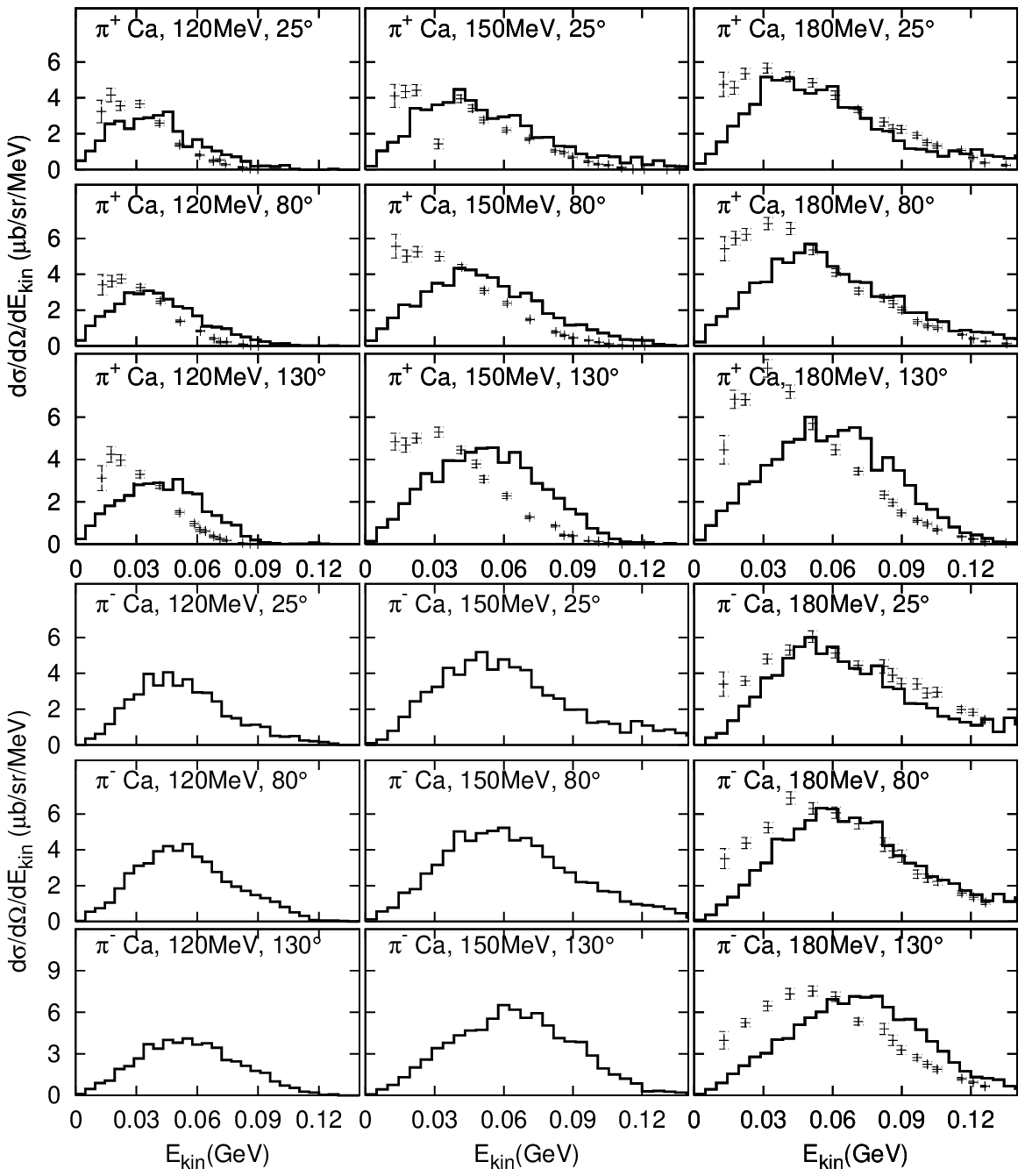}
\end{center}
\caption{Same as \fig{\ref{O_dOmegadE}} for $Ca$.}
\label{Ca_dOmegadE}
\end{figure}

\begin{figure}
\begin{center}
\includegraphics[]{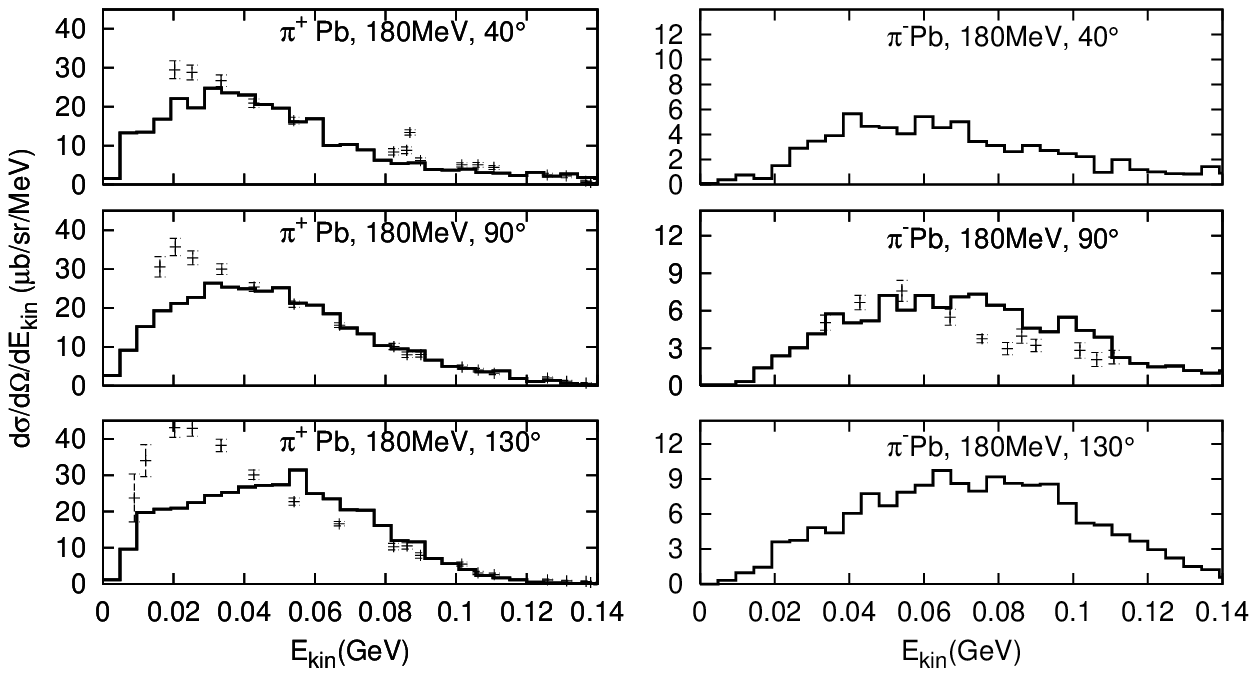}
\end{center}
\caption{Same as \fig{\ref{O_dOmegadE}} for $Pb$ at $\ekin=180\MeV$.}
\label{Pb_dOmegadE}
\end{figure}

Going into further details of the energy distribution of the produced pions, we show in figures \ref{O_dOmegadE}, \ref{Ca_dOmegadE} and \ref{Pb_dOmegadE} the results for $d\sigma / (d\Omega \, d\ekin) $ at different laboratory angles $\theta$, as a function of the kinetic energy of the outgoing pion $\ekin$. The overall agreement is good, better at forward and transverse angles than at backward angles. We observe a lack of pions with energies below $\ekin {\new \simeq} 30 \MeV$. This feature becomes more prominent when going from $O$ to $Pb$ and is present for both incoming $\pi^+$ and $\pi^-$. The same problem also shows up in the work of Vicente et al.~\cite{Vicente:1988iv}~(see their fig. 9). A solution to this problem is not clear. Due to the low-energy nature of those missing pions, one may speculate whether quantum mechanical effects are responsible for the enhancement, and therefore can not be described by a semiclassical transport theory.

\section{Summary}
We have studied pionic double charge exchange on different nuclear targets ($^{16}O$, $^{40}Ca$ and $^{208}Pb$) in the $\Delta$ region ($\ekin=120,150,180$~MeV) with a semiclassical couple channel transport model (GiBUU).

We have established the validity of the parallel ensemble scheme for this reaction, which is very sensitive to local density distributions by contrasting the results with those obtained in the more precise but time consuming  full ensemble method.

Furthermore, we compared the results of our model with the extensive set of data taken at LAMPF \cite{Wood:1992bi}, achieving a good agreement, not only for the total cross section, but also for angular distributions and double differential cross sections. Still, we miss some strength at backward angles and pion energies below $\ekin\approx 30\MeV$. The scaling of the total cross sections pointed out in \cite{Gram:1989qh} could be reproduced. However, we found that two important effects that break this scaling: neutron skins and Coulomb forces compensate each other.

We conclude that the implementation of pion rescattering and absorption in the GiBUU transport model successfully passes the demanding test of describing double charge exchange reactions. \newla{Thus the semi-classical approach is well suited to describe pion dynamics in nuclei for pion kinetic energies greater $\ekin\approx 30\MeV$.}
\begin{acknowledgments}
The authors thank Steven A. Wood for his cooperation by promptly  retrieving for us the experimental data.
OB is grateful to the rest of the GiBUU group, especially T. Falter, K. Gallmeister, T. Leitner and P. M\"uhlich for support and the good atmosphere and thanks S. Leupold for discussions. This work was supported by Deutsche Forschungsgemeinschaft (DFG). 
\end{acknowledgments}


\bibliography{literatur}

\end{document}